\documentclass[%
 reprint,
nofootinbib,
 amsmath,amssymb,
 aps,
]{revtex4-2}

\usepackage{graphicx}% Include figure files
\usepackage{dcolumn}% Align table columns on decimal point
\usepackage{bm}% bold math
\usepackage{aas_macros}%Journal macros
\usepackage{xcolor}

\newcommand{\vx}{\mathbf{x}}
\begin{document}

\preprint{APS/123-QED}

%\preprint{APS/123-QED}

\title{Searching for Parity Violation in SDSS DR16 Lyman-$\alpha$ Forest Data}

\author{Prakruth Adari}
\affiliation{Physics and Astronomy Department, Stony Brook University, Stony Brook, NY 11794}
% \email{prakruth.adari@stonybrook.edu}
\author{An\v{z}e Slosar}
\affiliation{Brookhaven National Laboratory, Upton NY 11973}

\date{\today}% It is always \today, today,

\begin{abstract}
The four-point correlation function is the lowest order correlation function for scalar fields that can be used to probe statistical parity invariance in an isotropic universe. 
There are intriguing claims of detection of parity violation in the 4-point function of BOSS galaxy clustering data. We apply the same estimator to the public SDSS Data Release 16 Lyman-$\alpha$ forest data. Lyman-$\alpha$ forest data probes a different redshift range and is sensitive to a different density regime using a completely different technique. A detection would therefore be a strong indication of new physics. We identify accurate covariance matrix as a crucial impediment to performing this measurement accurately, consistent with existing literature on galaxy 4-point function. We discuss several approaches to estimating the covariance matrix, several of which produce spurious detection. Using a robust, but very suboptimal, covariance matrix derived from subsample bootstrapping, we find no evidence for parity violation.
\end{abstract}

\maketitle

%\tableofcontents

\section{Introduction}

Testing parity invariance of the large scale structure of the universe is one of the foundational cosmological tests. 
% If the large-scale structure of the universe would statistically violate statistical parity, that would have profound impacts on our understanding of the fundamental physics. 
If the large-scale structure of the universe statistically violates parity, that would have a profound impact on our understanding of fundamental physics. 
It would imply that either the cosmological inflation contains parity-violating terms that generated parity violation in the initial conditions or, perhaps even more far-fetched, that gravitational evolution is parity violating, generating parity-violating distribution of matter in the universe from what were parity invariant initial conditions. 

Using scalar fields, the lowest order parity violating correlator is a 4-point function assuming an isotropic universe.
Geometrically, the action of parity transformation can be achieved using rotation in 3D for a line (2-pt) or a triangle (3-pt), but not for a general tetrahedron which corresponds to a 4-point function.
Recent searches for parity violations in galaxy clustering have found a detection in distribution of BOSS galaxies by two groups, albeit employing similar methods and the same data \cite{Philcox:2021hbm,Hou:2022wfj}.
When a similar search has been repeated for the CMB, on both temperature and polarization data, no signal has been found in either \cite{philcox2023cmb, Philcox:polarization}.
% When a similar search has been repeated on temperature distribution in the cosmic microwave background (CMB), no signal has been found \cite{philcox2023cmb}.
Under certain scaling assumption, the CMB result is much more constraining, but this statement hides considerable model dependence. It is therefore imperative to understand the original of this signal better and to search for it in other datasets.

In this paper, we repeat the same test on data from the Lyman-$\alpha$ forest data from eBOSS data\cite{SDSS:eboss}. Lyman-$\alpha$ forests trace the cosmic structure at different redshifts and with very different systematics compared to galaxy clustering. In this paper, we set to measure this quantity using the eBOSS Data Release 16 data \cite{SDSS:dr16}. We find no detection of parity violation, but we also again identify the covariance matrix of the measurement as the most difficult part of analysis, since the usual systematics present in the data do not generate parity violation signal unless one is already present. As a by-product, we have performed some of the first investigation of 3D polyspectra for Lyman-$\alpha$ forests on observed data.

\section{Data \& Method}

The main observable in the Lyman-$\alpha$ forest is the transmitted flux fluctuation (see \cite{Slosar_2011} for a pedagogical introduction):
\begin{equation}
    \delta_F = \frac{F(\vx)}{\bar{F}(\vx)}-1,
\end{equation}
The value of $\delta_F$ traces the underlying neutral hydrogen fluctuations in a very non-linear and mildly non-local manner. 
In absence of the large-scale photo-ionizing fluctuations, $\delta_F$ traces the linear matter fluctuations on the largest scales following linear bias models for the same reasons as galaxies do.
Most importantly, known physics cannot generate parity violating signal in $\delta_F$ unless these is some already present in the underlying fluctuations.

We use Sloan Digital Sky Survey's DR16  Lyman-$\alpha$ spectroscopic data \cite{SDSS:overview,SDSS:dr16,SDSS:eboss} as our input catalog. 
We rely on the public DR16 catalog\footnote{\url{https://www.sdss4.org/dr16/spectro/lyman-alpha-forest/}}. 
In this catalog, the quasars have already been continuum fitted and measurements were converted into wavelength estimates ($\lambda$), flux fluctuations ($\delta_F$), and flux fluctuation inverse variance weights ($w$). 
The wavelength is converted into redshift using $z = \lambda / \lambda_\alpha -1$, where $\lambda_\alpha=1215.7\,\AA$ is the rest-frame wavelength of the Lyman-$\alpha$ transition. Redshifts are in turn converted into comoving radial distances using a fiducial cosmological model with parameters $\Omega_c = 0.27$, $\Omega_b = 0.045$, $h=0.7$, $\sigma_8=0.8$ and $n_s = 0.96$. In total, we use 200,000 quasars and over $3.43 \times 10^{7}$ data points along the lines of sight. 
For pipeline testing we use a set of DR11 mock spectra \cite{lya_mocks} with similar number of quasars, total points along the line of sights, and redshift range.
The spectra are generated with the same cosmological parameters listed above.

We perform very modest data cuts, removing the lowest-SNR data. 
In particular, we require that $\lvert\delta_F\rvert < 5$ and $w > 0.1$ which leaves $3.38 \cdot 10^{7}$ points, removing $2\%$ of the data which is quite noisy. This cut is required by our approach at measuring the 4-point function as we describe next.

\begin{figure}
    \centering
    \includegraphics[width=\linewidth]{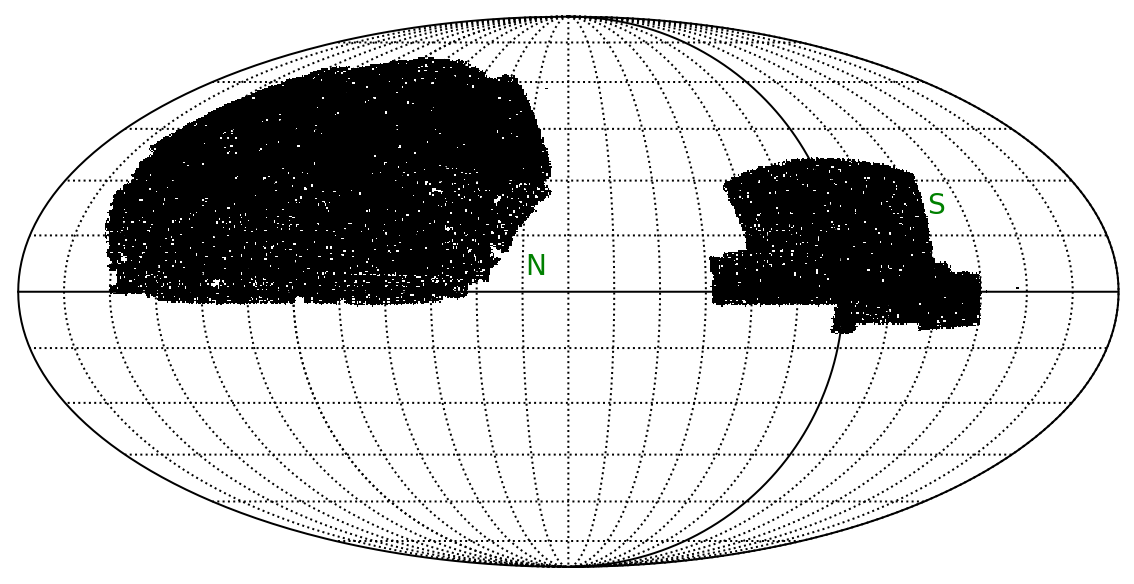}
    \caption{Distribution of quasars on sky. The left patch is labelled with ``N'' and the right is labelled ``S''.}
    \label{fig:quas_sky}
\end{figure}

To calculate the 4-point function, we employ a Landy-Szalay (LS) estimator as implemented in the  \verb|ENCORE| package \cite{Philcox:2021bwo}. LS estimator requires a data and random catalog, which encodes the survey window function.
We generate them as follows.
At every point for which we have a measurement of $\delta_F$, there is associated object in the LS catalogs with the same RA, DEC, and radial distance. 
The associated weights for the signal catalog are given by $w(1+\alpha \delta_F)$ while for the random catalogs they are given by $w$.
The factor $\alpha$ simply rescales the resulting 4-point function signal by $\alpha^4$. We have used $\alpha=0.1$ since \verb|ENCORE| requires, for purely technical reasons, that all weights are positive. We have tested that using a different $\alpha$ and applying appropriate rescaling does not change the result. 
We also apply a spatial cut in RA-DEC to split our data into 2 contiguous sets which can be seen in Fig. \ref{fig:quas_sky} separated by the galactic plane. The South (``S'')  set on the right contains 59,000 quasars with the remaining 151,000 quasars in the North (``N'') set on the left.

What we measure is the spherically averaged 4-point function, $\zeta$, which  is parameterized by 3 distances $r_1, r_2, r_3$ and the 3 angular momenta $l_1, l_2, l_3$. 
% The spherically averaged 4-point function, $\zeta$, is parameterized by 3 distances $r_1, r_2, r_3$ and the 3 angular momenta $l_1, l_2, l_3$. 
$\zeta$ can be expanded using a basis set $\mathcal{P}_{l_1, l_2, l_3}$ which transforms as $(-1)^{l_1 + l_2 + l_3}$ under the parity operation\cite{Cahn:2021ltp}.
Furthermore, it can be shown that due to the Wigner 3j terms and product of spherical harmonics (which are indexed by $l_1, l_2, l_3$), $\mathcal{P}_{l_1, l_2, l_3}$ is real when $l_1 + l_2 + l_3$ is even and imaginary otherwise \cite{Cahn:2021ltp}. 
The combination of these two properties along with the orthonormality of $\mathcal{P}_{l_1, l_2, l_3}$ offers a convenient deconstruction of $\zeta$ into a parity-even term, $\zeta_+$ which is a sum of real terms only, and a parity-odd term, $\zeta_-$ which is a sum of imaginary terms only\cite{PhysRevD.106.063501}.
Given a real density perturbation field we expect to measure $\zeta_- = 0$ and any deviation would be significant evidence of a parity violation. 
We use 10 radial bins of size $\Delta r = 15h^{-1}\textrm{Mpc}$ and set $l_i \leq 4$ which gives a total of 56 radial bins and 23 angular bins.
Further details on the 4-point function can be found in \cite{Philcox:2021bwo, Philcox:2021eeh, PhysRevD.106.063501, Cahn:2021ltp}.

Following previous works on parity violation\cite{philcox2023cmb}, we can calculate the number of linear modes that would be present in our dataset from any primordial signal\cite{Sailer:2021nmodes}.
We find there are  $2\times 10^5$ linear modes in the data making this comparable to the CMB results \cite{philcox2023cmb}.

\begin{figure}
    \includegraphics[width=\linewidth]{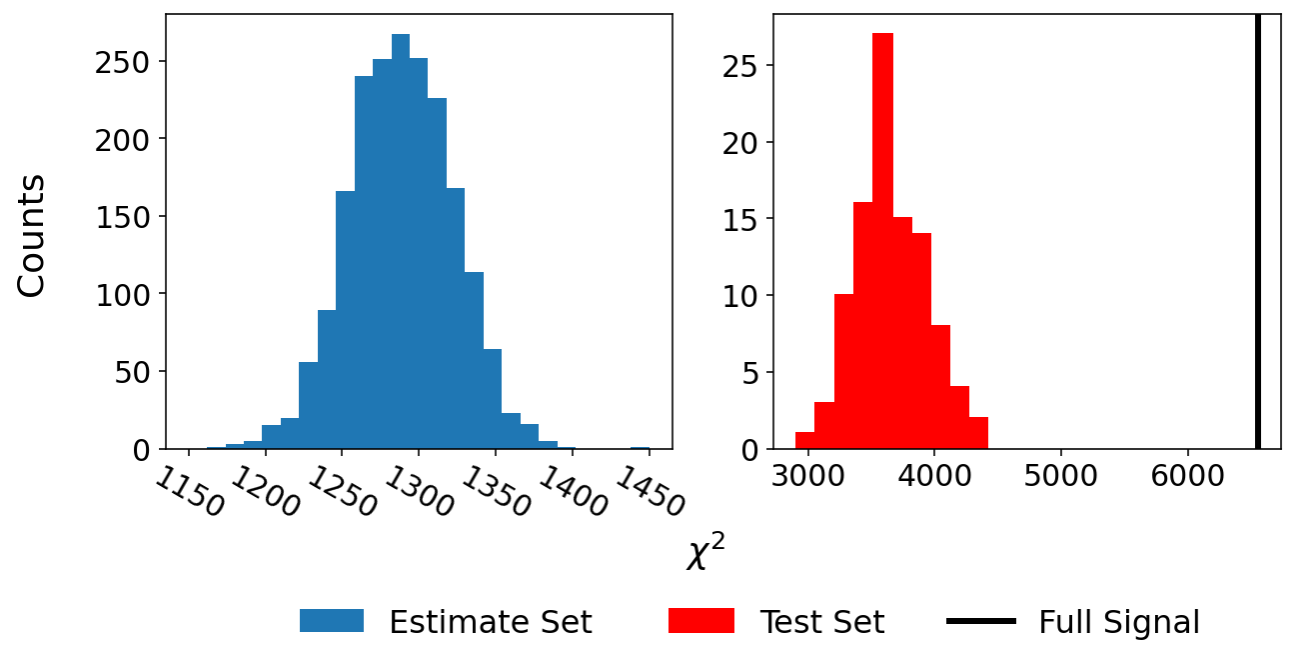}
    \caption{Histogram of $\chi^2$ scores when using the full covariance matrix $\Sigma_F$. The left panel plots the distribution of scores using the Estimate Set (ES) which are used to generate $\Sigma_F$ and the right is the distribution for the Test Set (TS). We also include the $\chi^2$ from our signal datavector on the right panel. Given that the TS comes from the same overall set of realizations, this mismatch in distributions is indicative of deeper problems with the covariance.}
    \label{fig:fullsigma_chi2}
\end{figure}

\begin{figure}
    \includegraphics[width=\linewidth]{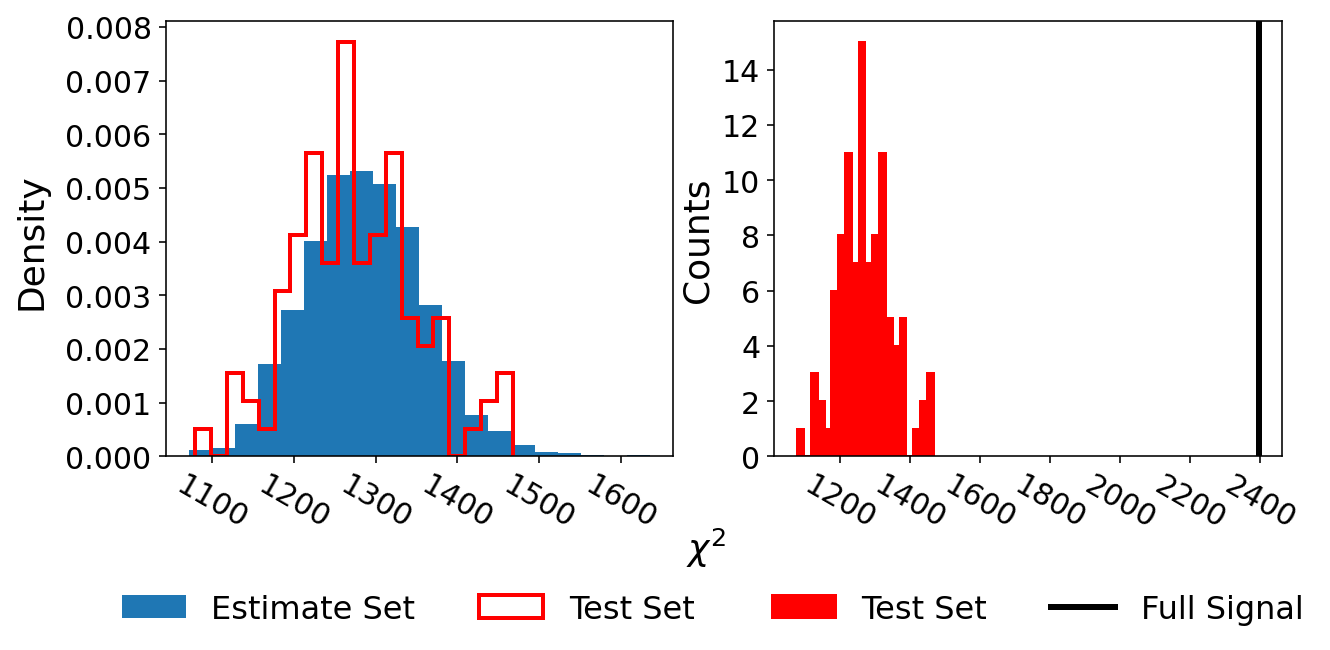}
    \caption{Similar to Fig~\ref{fig:fullsigma_chi2} but using $\Sigma_D$, the covariance matrix generated by taking the diagonal of $\Sigma_F$. To highlight the consistency between the Estimate Set, and Test Set, the two distributions are both shown on the left panel. The right panel plots the TS distribution along with $\chi^2$ from our signal datavector. While we do satisfy our consistency check, we find that this signal can be created on mock data as shown in Fig.~\ref{fig:mock_chi2_diagonal}. See Sec.~\ref{sec:cov_shuffle} for further details.}
    \label{fig:diagsigma_chi2}
\end{figure}

\begin{figure}
\includegraphics[width=\linewidth]{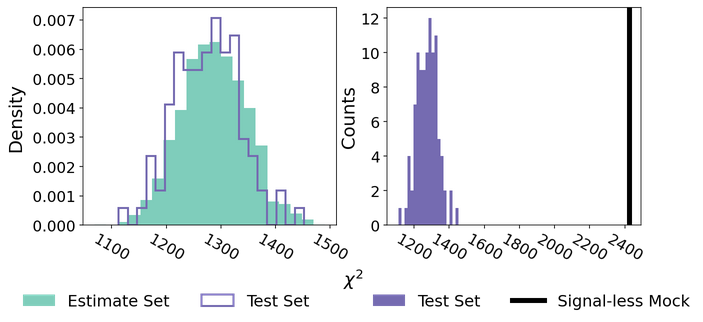}
\caption{Similar to Fig~\ref{fig:diagsigma_chi2} but using mock spectra. We create Mock Estimate Set (MES) and Mock Test Set (MTS) of sizes 1900 and 100 respectively. The panel on the left plots the distribution of $\chi^2$ scores of both the MES and MTS to highlight that we are satisfying our consistency check while the right panel plots the MTS scores and the mock signal $\chi^2$. Notably, there should be no signal on mock data.
% We use red and a hatch pattern to indicate that this plot was made using mock data instead of the signal data set.
}
\label{fig:mock_chi2_diagonal}
\end{figure}

\newcommand{\vd}{\mathbf{d}}
\subsection{Signal and Covariance Matrix}

We isolate the parity violating combinations of the 4-point function into a vector $\vd \in \mathbb{R}^{1288}$. In a parity conserving universe, the expectation value for all those 4 point correlators is exactly zero, and therefore the $\chi^2$ value defined as 
\begin{equation}\label{eq:chi2}
\chi^2 = \vd^T C^{-1}\vd
\end{equation}
is expected to be $\chi^2$ distribution. A large value of $\chi^2$ would be inconsistent with being drawn from this distribution and would signal detection of parity violation. 

The biggest problem is a robust estimation of the covariance matrix $C$. Since the variable of interest is a 4-point function of the underlying field, the estimator uncertainty, in general, depends on the 8-point function of the underlying field. The 8 point function can in turn be written to have contributions from four 2-point functions, followed by a product of two disconnected 3-point functions and 2-point functions, etc. all the way to the disconnected 8-point functions. This is in general extremely hard to compute, but given that individual field measurements are noise dominated, it is perhaps a valid assumption that the dominant contributions to uncertainty comes from products of 2-point function, which we well denote as 
\begin{equation}
    C^d_{ij} = \left< \delta_{F,i}  \delta_{F,j} \right>.
\end{equation}
This covariance matrix $C^d$ can in turn be divided into three regimes: i) diagonal part, i.e. variance of individual $\delta_F$ measurements is the largest one and it receives contributions from both noise and 1-dimensional correlations of the intrinsic signal along the line of sight; ii) the cross-covariance of pairs of $\delta_F$ measurements from the same quasar, it receives contributions from correlated noise due to continuum fitting and intrinsic signal along the line of sight; iii) the cross-covariance of pairs of $\delta_F$ measurements from pixels residing in nearby quasars receiving contribution only from the intrinsic signal. Pairs of $\delta_F$ measurements from distant quasars are uncorrelated.  When calculating the 2-point correlation function, it is usually a decent approximation to ignore the contribution from iii), since it is the smallest. 
\subsection{Covariance Matrix from Quasar Shuffling}\label{sec:cov_shuffle}

Based on this intuition, our first attempt to calculate covariance matrix was to randomly shuffle quasar positions. We swapped the RA and DEC for all points along a line of sight (LOS) with another LOS in order to maintain the covariance terms from i and ii, and then the corresponding 4-point function was calculated.
This process was repeated 2000 times.
1900 were used to estimate the covariance, $\Sigma_F$, and the remaining 100 were preserved as a self-consistency check.
The former is labelled the Estimate Set (ES) and the latter as the Test Set (TS).
The distribution of $\chi^2$ scores on the ES is shown in the left panel of Fig.~\ref{fig:fullsigma_chi2} and the TS is shown on the right. 
As seen in the right panel, the distribution is massively misaligned with the expected mean of 1288. 
We remedy this by taking only the diagonal terms in $\Sigma_F$, $\Sigma_D$, which produces the $\chi^2$ distributions shown in Fig.~\ref{fig:diagsigma_chi2}.
As shown on the panel on the left, there is good agreement between the ES and TS $\chi^2$ distributions.
The mean and standard deviation for each set and variation (see Appendix~\ref{appendix:covariance}) can be found in Table~\ref{tab:covariances}.
% However, as seen in the right panel of Fig.~\ref{fig:fullsigma_chi2} which plots the histogram of the $\chi^2$ scores  or from the mean in Table~\ref{tab:covariances}, the consistency check fails with a mean $\chi^2 \approx 3600$.  We remedy this by taking only the diagonal terms which does satisfy the self-consistency check as seen in Fig.~\ref{fig:diagsigma_chi2} and Table~\ref{tab:covariances}. 

We see that in both cases, the real, unshuffled data is inconsistent with the test data with a very large significance. 
This is a completely internally derived covariance matrix that, especially in the diagonal case, seems to pass our consistency test.
However, when our entire process is repeated on a single DR11 mock -- shuffling RA\&DEC, calculating $\xi$, estimating covariance, and taking the diagonal term -- we find a spurious signal from our mock which is shown in Fig.~\ref{fig:mock_chi2_diagonal}.
This implies that shuffling is a fundamentally flawed as it is clearly destroying correlations that contribute significantly to the measurement covariance matrix.  We have attempted a few other approaches to estimating the covariance matrix without success -- for completeness we detail them in then Appendix~\ref{appendix:covariance}.

\begin{figure}
\includegraphics[width=\linewidth]{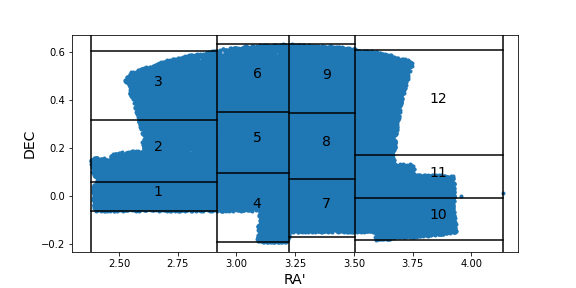}
\caption{Splitting the S patch into 12 subpatches each with 4882 quasars per subpatch. The x-axis corresponds to a shifted RA to have a contiguous range.}
\label{fig:patch_sky_12}
\end{figure}

\begin{figure}
\includegraphics[width=\linewidth]{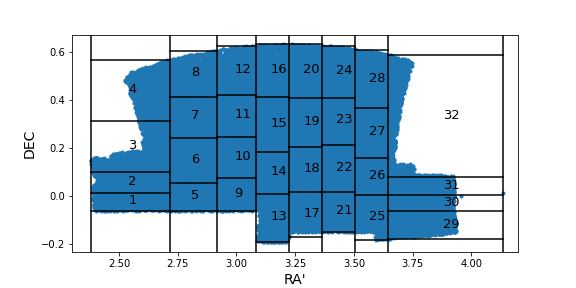}
\caption{Same as Fig~\ref{fig:patch_sky_12} but with 32 subpatches each with 1831 quasars per subpatch.}
\label{fig:patch_sky_32}
\end{figure}

\begin{figure*}
\begin{center}
\includegraphics[width=\linewidth]{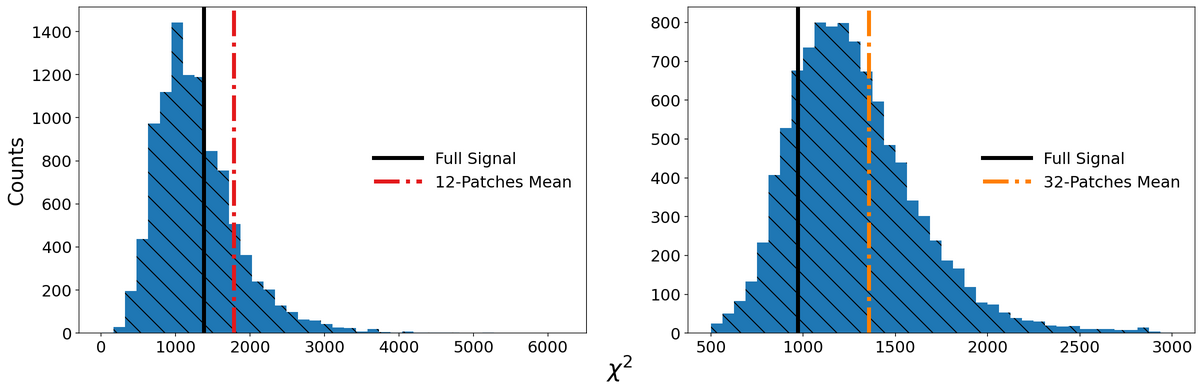}
\caption{Distribution of $\chi^2$ scores using the combined diagonal bootstrapped covariance matrix. Left plot uses 12 subpatches while right uses 32 subpatches. The red and black lines correspond to the $\chi^2$ score using the full data set or the mean of the subpatches correspondingly.}
\label{fig:bootstrap_chi2}
\end{center}
\end{figure*}

% \begin{figure*}
%     \begin{center}
%         \includegraphics[width=\linewidth]{patches_diag_cov_N.png}
%         \caption{Same as Fig.~\ref{fig:bootstrap_chi2} but on the Northern patch which we distinguish using a hatched pattern.}
%         \label{fig:bootstrap_chi2N}
%     \end{center}
% \end{figure*}

\subsection{Covariance Matrix from Bootstrap Resampling}
The spurious detection generated from quasar shuffling motivates our final attempt which uses a bootstrap resampling method. 
Each patch (N or S) is split into $N_p=$12 and $N_p=32$ subpatches as seen in Fig~\ref{fig:patch_sky_12} and Fig~\ref{fig:patch_sky_32} for the S patch. 
We calculate the datavector in each subpatch independently and use the standard bootstrap resampling algorithm to calculate the corresponding covariance.

This means we repeatedly draw $N_p$ values from the subpatches \emph{with replacement} and then calculate the mean. This is repeated 10,000 times and that sample is used to estimate a covariance which we assume to be diagonal due to the relatively small number of subpatches.

% We assume that the covariance for the Northern patch can be related to the South with a simple scaling of the number of quasars.
The diagonal bootstrapped covariances, $C^d_N$ and $C^d_S$, from the N and S patch respectively, are then combined using \begin{equation}\label{eq:cov_add}
C_{NS}^d = \left[ \left(C_N^d\right)^{-1} + \left(C_S^d\right)^{-1} \right]^{-1}.
\end{equation}
Similarly the corresponding signal vector, $\vd_{NS}$, from $\vd_N$ and $\vd_S$ which come from the mean over all the patches, via \begin{equation}\label{eq:sig_add}
\vd_{NS} = C_{NS}^d \left[ \left(C_N^d\right)^{-1}\vd_N + \left(C_S^d\right)^{-1}\vd_S \right].
\end{equation}
Results for both values of $N_p$ are shown in Fig~\ref{fig:bootstrap_chi2} where the data results are consistent with no detection of parity-violating 4-point function.

\section{Conclusions}

We have attempted to test for the presence of a parity-violating 4-point function in the Lyman-$\alpha$ forest data. While the measurements can be calculated using several existing and publicly available codes, determining a reliable covariance matrix is very difficult.

We have attempted to determine the covariance matrix using quasar shuffling, which preserves the dominant correlations present in a single quasar. The diagonal version of this covariance matrix passes the basic sanity check in that it produces a viable distribution of $\chi^2$ on a test shuffling set. While it produces a strong detection, the same is obtained when the procedure is repeated on a mock dataset. This implies that the basic premise of quasar-shuffling approach, namely that the intra-quasars correlations dominate the uncertainty is wrong.
 
% This is particularly important as recent work has suggested that the detection in the BOSS survey can arise from underestimating the covariance \cite{Philcox:covariance}

We have not attempted to measure the covariance matrix using mocks alone for two reasons. First, they correspond to an older, smaller catalog (DR11 rather than DR16) and more importantly, are unlikely to capture the higher-order correlations correctly since these were never designed to be used beyond 2-point function measurements. Recent work which suggests that the BOSS detection can be attributed to an misestimated covariance \cite{Philcox:covariance} motivates our conservative and suboptimal approach which produced a null result.

% Using a bootstrap resampling method we did find a suitable covariance matrix that produced a null result. However, it is clear that this result is very suboptimal. 
As shown in \cite{fuma}, the distribution of $\chi^2$ can be made to have the correct mean by rescaling, but its spread around this mean for mock data "measures" the deviation of the covariance matrix from the true one.  Given our number of points, the spread in $\chi^2$ should be around 50, but it is at least an order of magnitude larger as shown in Figure \ref{fig:bootstrap_chi2}. This means that it is possible that even in this dataset, the signal could be present but we are unable to ascertain its existence due to overly noisy covariance matrix.

The question remains on what would be the procedure to obtain a useful covariance matrix. As outlined above, at the level of disconnected covariance matrix can in principle be calculated using power-spectrum alone. For this to work, one would need to integrated over all possible combinations of 8 pixels in the data and for each set of 8 over all possible pairs. This is computationally prohibitive, even on a small subset of data. A simpler combination would be to assume survey to be completely homogeneous and then calculate the relevant 8-point function in Fourier space for a fully homogeneous box, followed by approximate scaling to the realistic survey. This would involve significant work and while feasible, it is not clear that it would be accurate enough. An alternative approach would be to compute the 4-point function as a cross-correlation of pairs of field, each of which is in turn a product of filtered raw maps. This has the advantage that the technology for calculating cross-power spectra is well developed and that the auto-power spectra of squared fields already contains contributions from connected terms (see \cite{2024arXiv240907980H} for in-depth application of this method to bi-spectrum). Unfortunately, all of these methods would require significant investment of resources into computing the covariance matrix, which might or might not work. Fundamentally, the problem lies in the fact that exact signature of parity violation is not theoretically supported -- if we could compress the parity violation into a single number (as opposed to a vector with over a thousand degrees of freedom), the error on such number could be reliably measured from the bootstrapping approach used here. 

In conclusion, we show that the test of parity violation based on 4-point function of a scalar field is quite challenging in practice due to difficulty in obtaining a reliable covariance matrix.

\section*{Acknowledgements}

We acknowledge useful discussions with Oliver Philcox.
We acknowledge use of the Core Cosmology Library \cite{pyccl}, Matplotlib \cite{Hunter:2007}, Numpy \cite{harris2020array}, Scipy \cite{2020SciPy-NMeth}, Healpy \cite{healpy1, healpy2}, and the ColorBrewer.

\appendix

\section{Covariance Matrix Estimations}
\label{appendix:covariance}
We tried a number of methods to estimate the covariance matrix via quasar shuffling by varying terms included.
As a brief reminder, our data had been binned into 23 angular bins and 56 radial bins leading 1288 points.
We test using five subsets of $\Sigma_F$:
\begin{enumerate}
    \item \textbf{Full} - Use all radial and angular bins. 
    \item \textbf{Diagonal} - The simplest case wherein we remove any non-diagonal terms. Each row has 1 non-zero value.
    \item \textbf{Same Radial Bin} - All values from different radial bins are set to 0. Each row has 23 non-zero values. 
    \item \textbf{Neighboring Radial Bin} - All values that are not in the same or neighboring radial bin are set to 0. The smallest and largest radial bins have 46 non-zero values while all middle radial bins have 69 non-zero values.
    \item \textbf{Same Angular Bin} - All values from different angular bins are set to 0. Each row has 56 non-zero values.
\end{enumerate}
For all the methods, we generate 2000 realizations by shuffling quasars in the S patch, use 1900 to estimate the covariance (ES) and 100 as a consistency check (TS) as in the main text.
We have checked that doing a completely random shuffling and shuffling within redshift bins leads to similar $\chi^2$ distributions.

The mean and 1$\sigma$ standard deviation for the various methods on the ES and the TS can be found in Table.~\ref{tab:covariances}.
We found that the purely diagonal covariance, $\Sigma_D$, led to the most consistent $\chi^2$ between the TS and ES (Fig~\ref{fig:diagsigma_chi2}).
As discussed in the main text, this consistency check is not sufficient as the mocks produced a spurious signal as seen in Fig.~\ref{fig:mock_chi2_diagonal}.

\begin{table*}
    \centering
 \begin{tabular}{c|c c c c c}
     & Full & Diagonal & Same Radial Bin & Neighboring Radial Bin & Same Angular Bin \\
     \hline
    Estimate Set & 1288 $\pm$ 43.3 & 1288 $\pm$ 72.1 & 1288 $\pm$ 70.9 & 1296 $\pm$ 68.4 & 1288 $\pm$ 68.4 \\ 
    Test Subset & 3659 $\pm$ 291.4 & 1296 $\pm$ 75.3 & 1309 $\pm$ 76.6 & 1351 $\pm$ 75.5 & 1332 $\pm$ 74.9 \\
\end{tabular}
    \caption{Mean and standard deviation of $\chi^2$ distributions on simulation data using a variety of covariance matrices.}
    \label{tab:covariances}
\end{table*}

% \nocite{*}

\bibliography{apssamp}% Produces the bibliography via BibTeX.

\end{document}